# Nonmagnetic invisibility cloak with the minimized scattering cross section

Lujun Huang[1], Daming Zhou[1], Jian Wang[1], Zhifeng Li[1], Xiaoshuang Chen[1,*], and Wei Lu[1]

[1] *National Laboratory for Infrared Physics, Shanghai Institute of Technical Physics, Chinese Academy of Science, Shanghai 200083, People's Republic of China*

We propose one kind of transformation functions for nonmagnetic invisibility cloak with minimized scattering on the basis of generalized transformation. By matching the impedance at the outer surface of the cloak, the transformations with two parameters are determined. To confirm the performance of the cloak, full wave simulation based on the finite element method is carried out. Furthermore, total scattering cross section is computed to better illustrate the scattering characteristics of cloak with different parameters. In addition, based on the effective media theory, alternating layered system composed of two isotropic materials is employed to realize the cloak practically.

Corresponding Author:*Email: xschen@mail.sitp.ac.cn

## I. INTRODUCTION

Recently, the increasing interest has been attracted on designing the invisibility cloak, [1-31] which was first proposed by Pendry and Leonhardt in 2006, respectively. [1-2] An ideal cloak can guide the electromagnetic wave flowing smoothly around the object and render the object inside it invisible to the outside observers. Inspired by the idea of the invisibility cloak, the first cylindrical cloak with simplified material parameters was soon verified at the microwave frequency by Schruig et.al. [3-4] However, the method of designing the microwave cloak cannot be directly applied to the construction of optical cloaks because materials with the changing permeability are difficult to be realized. [26] To achieve invisibility cloak at optical frequency, Cai et.al proposed a nonmagnetic cloak with simplified material parameters. [5] However, such a cloak has a drawback that significant scattering outside the cloak exists due to the mismatched impedance at the outer surface. In order to minimize the scattering of the cloak, various transformation functions, such as quadratic and piece-wise high order functions, were proposed. [27-31]

Recently, we proposed a generalized transformation to realize the nonmagnetic cloak with minimized scattering. [31] It is found that the scattering cross section reduces significantly as long as the transformation function satisfies certain conditions. In this paper, on the basis of the generalized transformation, one kind of transformation functions with two parameters is proposed. By changing these two different parameters, the nonmagnetic invisibility cloak with the minimized scattering is determined. Numerical simulation based on the finite element method is carried out to confirm the cloak's performance. Furthermore, total scattering cross section for different cloaks is calculated to compare their scattering properties. In addition, based on the effective medium theory, alternating layered system with two isotropic materials is applied to realize such a cloak practically.

## II. SCATTERING THEORY OF THE INVISIBILITY CLOAK

Let us first consider a two dimensional cylindrical invisibility cloak and restrict our analysis on the normal incidence. For transverse-magnetic (TM) incidence, following the procedure illustrated in Ref,[1] the ideal cylindrical cloak can be easily achieved by squeezing a cylindrical air region $r'\leq b$ into the cylindrical annular region $a\leq r\leq b$. Assuming the coordinate transformation function $r'=f(r)$, $\theta'=\theta$, $z'=z$, which satisfies $f(a)=0$ and $f(b)=b$. The relative permittivity and permeability of the ideal cloak can be calculated by the Jacobian matrix

$$\varepsilon_r = f(r)/[rf'(r)]$$
$$\varepsilon_\theta = rf'(r)/f(r)$$
$$\mu_z = f(r)f'(r)/r \tag{1}$$

where $f'(r) = df(r)/dr$.

Within the cloak shell, the general wave equation that governs the $H_z$ field can be written as [8]

$$\frac{1}{\mu_z}\frac{1}{r}\frac{\partial}{\partial r}(\frac{r}{\varepsilon_\theta}\frac{\partial H_z}{\partial r}) + \frac{1}{\mu_z}\frac{1}{r^2}\frac{\partial}{\partial \theta}(\frac{1}{\varepsilon_r}\frac{\partial H_z}{\partial \theta}) + k_0^2 H_z = 0 \tag{2}$$

where $k_0$ is the wave vector of the EM wave in vacuum. The time harmonic mode $\exp(i\omega t)$ has been applied in deriving Eq. (2). By employing the variable separation method $H_z(r,\theta) = \Psi(r)\Theta(\theta)$, Eq. (2) can be decomposed into two separation equations. The solution for angle-dependent equation is $\Theta(\theta) = \exp(im\theta)$, where m is an integer. Another radial dependent equation is written as

$$\frac{d}{dr}(\frac{r}{\varepsilon_\theta}\frac{d\Psi}{dr}) + k_0^2 r\mu_z \Psi - \frac{m^2}{r\varepsilon_r}\Psi = 0 \tag{3}$$

The EM fields outside the cloak are well known and can be expressed with the cylindrical functions. Based on the above analysis, the magnetic fields in each layer are formulated as follows,

$$H_z(r,\theta) = \begin{cases} 0 & a < r \\ \sum_m (\alpha_m^i J_m(k_0 f(r)) + \alpha_m^s H_m^{(1)}(k_0 f(r))) \exp(im\theta) & a < r < b \\ \sum_m (\beta_m^i J_m(k_0 r) + \beta_m^s H_m^{(1)}(k_0 r)) \exp(im\theta) & r > b \end{cases} \quad (4)$$

where $J_m$ and $H_m^{(1)}$ are mth-order Bessel function and Hankel function of the first kind, respectively. Considering the continuity of $H_z$ and $E_\theta = \frac{1}{i\omega\varepsilon}\frac{\partial H_z}{\partial r}$ at the boundary of $r=a$ and $r=b$, the relationship between the unknown coefficients can be expressed as

$$\alpha_m^i J_m(k_0 f(a)) + \alpha_m^s H_m^{(1)}(k_0 f(a)) = 0 \quad (5a)$$

$$\frac{f(a)}{a}[\alpha_m^i J_m'(k_0 f(a)) + \alpha_m^s H_m^{(1)'}(k_0 f(a))] = 0 \quad (5b)$$

$$\alpha_m^i J_m(k_0 f(b)) + \alpha_m^s H_m^{(1)}(k_0 f(b)) = \beta_m^i J_m(k_0 b) + \beta_m^s H_m^{(1)}(k_0 b) \quad (5c)$$

$$\frac{f(b)}{b}[\alpha_m^i J_m'(k_0 f(b)) + \alpha_m^s H_m^{(1)'}(k_0 f(b))] = \beta_m^i J_m'(k_0 b) + \beta_m^s H_m^{(1)'}(k_0 b) \quad (5d)$$

From the above equation, one can get the coefficients for the magnetic field,

$$\alpha_m^s = \beta_m^s = 0, \alpha_m^i = \beta_m^i \quad (6)$$

Obviously, the scattering fields within the cloak medium and outside the cloak are zero. That demonstrates the function of the perfect invisibility cloak.

However, it is very difficult to fabricate such an ideal cloak practically since the relative permittivity and permeability become infinite at the inner boundary $r=a$. Moreover, owing to the saturation of the magnetic response of the split-ring resonator (SRR) at the optical frequency,[26] constructing an ideal cloak with the changing permeability becomes very hard and the loss of SRR also becomes significant at the optical frequency. To circumvent these difficulties, Cai et.al proposed a nonmagnetic cloak by setting relative permeability as unity.[5] Nevertheless, such a cloak has significant scattering due to the mismatched impedance at the outer boundary r=b, which greatly degrades the performance of the invisibility cloak. In order to minimize the scattering of the cloak, several nonmagnetic cloaks with different transformation function has been proposed.[27-30] Based on the generalized transformation, it is known that a nonmagnetic cloak with the minimized scattering can be realized as long as the transformation function satisfies some certain conditions [31]. Still considering transformation function $r'=f(r)$, $\theta'=\theta$, $z'=z$, by setting relative permeability equal unity and keeping $\varepsilon_r\mu_z$ and $\varepsilon_\theta\mu_z$ unchanged, the wave trajectory in the cloak medium can be maintained, and thus the relative permittivity of the cloak shell can be expressed as

$$\varepsilon_r = (f(r)/r)^2, \varepsilon_\theta = (f'(r))^2, \mu_z = 1 \tag{7}$$

In order to achieve the nonmagnetic cloak with minimized scattering, the transformation function should satisfy

$$f(a)=0, f(b)=b, Z_{r=b} = \sqrt{\mu_z/\varepsilon_\theta} = 1/f'(b) = 1 \tag{8}$$

Then the transformation function can be expressed in the following form

$$f(r) = g(r)\sum_n a_n(r-a)^n$$
$$g(b)\sum_n a_n(b-a)^n = b$$
$$g(b)\sum_n a_n n(b-a)^{n-1} + g'(b)\sum_n a_n(b-a)^n = 1 \tag{9}$$

where n is an arbitrary positive number and g(r) is a continuous function which does not contain the term (r-a).

Substituting Eq. (7) into Eq. (3), Eq. (3) can be rewritten as

$$\frac{d}{dr}(\frac{r}{(f'(r))^2}\frac{d\Psi}{dr}) + k_0^2 r\Psi - \frac{m^2 r}{(f(r))^2}\Psi = 0 \tag{10}$$

Similarly, the EM fields in three different regions can be written as follows

$$H_z(r,\theta) = \begin{cases} 0 & r<a \\ \sum_m (\alpha_m^i Q_m(k_0 r) + \alpha_m^s R_m(k_0 r))\exp(im\theta) & a<r<b \\ \sum_m (\beta_m^i J_m(k_0 r) + \beta_m^s H_m^{(1)}(k_0 r))\exp(im\theta) & r>b \end{cases} \tag{11}$$

where the transcendental functions $Q_m$ and $R_m$ are the solutions to the Eq. (10) in the cloak shell; $J_m$ and $H_m^{(1)}$ are the mth-order Bessel function and Hankel function of the first kind. The coefficients for magnetic fields can be derived by applying the continuous conditions of EM fields at the two interface $r=a$ and $r=b$, which is

expressed as

$$\alpha_m^i Q_m(k_0 a) + \alpha_m^s R_m(k_0 a) = 0 \tag{12a}$$

$$\frac{1}{f'^2(a)}[\alpha_m^i Q_m(k_0 a) + \alpha_m^s R_m(k_0 a)] = 0 \tag{12b}$$

$$\alpha_m^i Q_m(k_0 b) + \alpha_m^s R_m(k_0 b) = \beta_m^i J_m(k_0 b) + \beta_m^s H_m^{(1)}(k_0 b) \tag{12c}$$

$$\frac{1}{f'^2(b)}[\alpha_m^i Q_m'(k_0 b) + \alpha_m^s R_m'(k_0 b)] = \beta_m^i J_m'(k_0 b) + \beta_m^s H_m^{(1)}{}'(k_0 b) \tag{12d}$$

Since functions $Q_m$ and $R_m$ are always different from the Bessel function and Hankel function, both the coefficients $\alpha_m^s$ and $\beta_m^s$ are nonzero. Therefore such a nonmagnetic cloak inevitably has scattering.

## III. SIMULATION RESULTS AND DISCUSSIONS

Considering a transformation function which has the following form

$$\begin{aligned} r' &= f(r) = (mr^s + p)(r-a)^t \\ m &= \frac{(b-a-bt)}{(b-a)^{t+1} b^{s-1} s} \\ p &= \frac{sb(b-a) - b(b-a-bt)}{(b-a)^{t+1} s} \end{aligned} \tag{13}$$

where both the sign of parameters s and t are positive. It is not difficult to demonstrate that the above transformation satisfy Eq. (8).

The material parameters of the cloaking medium can be obtained through substituting Eq. (13) into Eq. (7),

$$\begin{aligned} \varepsilon_r &= ((mr^s + p)(r-a)^t / r)^2 \\ \varepsilon_\theta &= [msr^{s-1}(r-a)^t + t(mr^s + p)(r-a)^{t-1}]^2 \\ \mu_z &= 1 \end{aligned} \tag{14}$$

Different from the quadratic coordinate function, such a nonmagnetic cloak has no size restriction since the transformation function is monotonically increasing at $a \leq r \leq b$. The monotonic property can be demonstrated by $f'(r) = msr^{s-1}(r-a)^t + t(mr^s + p)(r-a)^{t-1} > 0$. Moreover, the cloak has no unphysical singularity in the inner boundary while t is greater than 1. For $0<t<1$, the azimuthal permittivity approaches infinity at the inner boundary r=a, thus more attention should be paid to while such a cloak is contructed practically. Figure. 1(a) and (b) show the material parameters of the cloak shell with different s and t while the outer radius b is twice of the inner radius a. It is shown that the relative permittivity equals unity at the outer boundary r=b, which indicates that the impedance matches at the outer boundary, and thus to a great extent reduces the scattering outside the cloak. As shown in Figure. 1(a), when s is kept as constant 1, the radial permittivity $\varepsilon_r$ increases slowly from 0 to 1. Nevertheless, the azimuthal permittivity $\varepsilon_\theta$ has a total different situation. With the increasing radius $r$, $\varepsilon_\theta$ decreases sharply from infinite to unity for t=0.5 while decreases relatively slowly from 8.9 to 1 for t=1. In contrast, $\varepsilon_\theta$ first increases from 0 to 6.4, and then decreases from 6.4 to 1. Figure. 1(b) presents the relationship between material parameters and radius while t is 1. It is found that relative permittivities $\varepsilon_\theta$ and $\varepsilon_r$ vary little even when s changes from 0.5 to 1.5.

In order to confirm the performance of the cloaks with different parameters s and t, full-wave simulation based on the finite element method (FEM) is performed. Figure. 2 shows the computational domain for a two dimensional full-wave simulation. Perfect Matched layer (PML) is set as the boundary condition of the simulation region. A TM polarized uniform plane wave (wavelength λ=a) incident on a thin perfect magnetic conductor shell (PMC) of radius a surrounded by a cloak shell with the outer radius b=2a. The material parameters of the cloak shell are given by Eq. (14). In the following part, we discuss the effects of the parameters s and t in the transformation function on the cloak's performance.

(a) Let us first set s=1, then the transformation function can be further simplified as

$$r' = f(r) = [\frac{(b-a-bt)r}{(b-a)^{t+1}} + \frac{b(b-a)-b(b-a-bt)}{(b-a)^{t+1}}](r-a)^t \tag{15}$$

The relative permittivity of the cloak medium can be obtained by substituting the above equation into Eq. (7). Clearly, the transformation function degenerates into the case in Ref. [28] for t=1. Due to the matched impedance at the outer boundary r=b, the scattering field of the cloak is greatly reduced compared to the nonmagnetic cloak with the simplified parameters in Ref. [5]. To study the function of the nonmagnetic cloak with different t, the value of t ranging from 0 to 3 is considered. The results of magnetic field distribution for the cloak are shown in Figure. 3(a) and (b) while t is equal to 0.5 and 1.5, respectively. It can be seen that the magnetic fields outside the cloak are almost undisturbed as if no object locates at the coordinate origin. Inside the cloak shell, the magnetic fields are clearly excluded from the interior PMC object. To compare the scattering of these two cloaks, the scattering fields outside the cloak are also calculated and plotted in Figure 3(c)-(d). It can be observed that the scattering fields range from -0.41 to 0.36 while t is set as 0.5. Nevertheless, for the case of t=2, the scattering field outside the cloak increase significantly and has a range from -0.783 to 0.685. Thus, it can be inferred that the scattering cross section is very sensitive to the value of t even though the impedance at the exterior boundary matches perfectly with the free space. To better illustrate the scattering characteristics of the cloaks, total scattering cross section (SCS) of the cloak with different parameters is calculated. For the sake of straightforward, the total SCS for the different cloak is normalized to the counterpart of the PMC. Figure. 4 shows the normalized SCS of the cloak while the value of t varies from 0.1 to 3.0. As shown in figure, it is observed that the normalized SCS decreases significantly when t increases from 0.1 to 0.8. The normalized SCS has a minimum value 0.12 when t is 0.8. If we keep increasing the value of t from 0.8 to 3, the normalized SCS increases from 0.12 to 0.44. Moreover, the normalized SCS is up to 1.1 when t increases to 8. That also means that the nonmagnetic cloak with a larger t no longer makes the object inside it invisible. Thus, from the above analysis, this kind of cloak has the minimum SCS for t=0.8. Nevertheless, we should note that the azimuthal permittivity becomes infinite at the inner interface r=a while such a cloak is implemented practically. As a consequence, from the prospective of practical realization, t should be chosen as unity.

(b) Let's proceed to fix t=1 and consider the influence of s to the scattering of cloak. According to the Eq. (13), we can express the transformation function as

$$r' = f(r) = [-\frac{ar^s}{(b-a)^2 b^{s-1} s} + \frac{sb^2 + (1-s)ab}{(b-a)^2 s}](r-a) \tag{16}$$

The material property of the cloak shell is determined by substituting the above equation into Eq. (7). Like the cloak proposed in (a), such cloak also has no restraint on the thickness of the cloak. Moreover, this kind of nonmagnetic cloak has another merit that no unphysical singularity exist at the inner boundary r=a. To demonstrate the performance of the cloaks with different s, the spatial distribution of the magnetic fields is plotten in Figure. 4(a) and (b). Obviously, all of the cloaks perform well in guiding the electromagnetic wave smoothly around the exterior surface of PMC without perturbing the exterior fields. After passing the cloaking area, the light rays restore their original direction as if the space is empty. Figure. 5(c) and (d) also present the scattering fields outside the cloak shell. It can be found that the scattering fields outside the cloak for s=0.5 are nearly the same as that of s=2. Figure. 6 presents the normalized SCS of the cloaks with different s. From the figure, we find that the normalized SCS increase very slowly even though s varies from 0.1 to 3.0. In other words, the variation of s has little effect to the total SCS of the cloak.

(c) For s≠1 and t≠1, the normalized SCS of the cloak is computed and plotted in Figure. 7 for 0<s≤2 and 0<t≤2. Clearly, all SCS ascend very slowly while t keeps unchanged except for the case of t=0.2. This again demonstrates that the value of t has little influence on the total SCS of the cloak. Moreover, for the same s, the normalized SCS has a minimum value less than 0.12 as t varies from 0.2 to 2.0.

Next, we proceed to address the problem on how to realize the nonmagnetic invisibility cloak with the isotropic and homogeneous material. It is required that the material parameters of the nonmagnetic invisible cloak are anisotropic and radius-dependent. To realize such a cloak practically, an alternative layered system based on the effective medium theory is employed.[20-22] Such a system has a merit that the material in each layer is homogeneous and isotropic. Here, we first divide the cloak shell into N concentric layers composed of alternative layers A ($\varepsilon=\varepsilon_A$) and B ($\varepsilon=\varepsilon_B$). When the thickness of the layer is much less than the wavelength (long wavelength limit) an anisotropic effective medium can be realized. For the sake of simplicity, we assume that the thickness of isotropic material A is the same as that of B. According to the effective medium theory, the anisotropic material parameters can be formulated as

$$\overline{\varepsilon} = \begin{bmatrix} \varepsilon_r & 0 \\ 0 & \varepsilon_\theta \end{bmatrix} = \begin{bmatrix} \dfrac{2\varepsilon_A \varepsilon_B}{\varepsilon_A + \varepsilon_B} & 0 \\ 0 & \dfrac{\varepsilon_A + \varepsilon_B}{2} \end{bmatrix}$$

(17)

Then the permittivies of A and B can be expressed as follows,

$$\varepsilon_A = \varepsilon_\theta + \sqrt{\varepsilon_\theta^2 - \varepsilon_r \varepsilon_\theta}$$
$$\varepsilon_B = \varepsilon_\theta - \sqrt{\varepsilon_\theta^2 - \varepsilon_r \varepsilon_\theta}$$

(18)

Thus, the material parameters of A and B in each layer are obtained.

While the cloak shell is divided into 10 layers with the identical thickness a/10, the material parameters for A and B can be calculated accordingly by combining Eq. (14) and Eq. (18). With the increase of the radius, the relative permittivity of dielectric A decrease gradually from 16.81 to 1.72 while the permittivity of dielectric B increase from 0.01 to 0.70. To study the effect of N to the performance of the cloak, the cloak shell with 20 layers is also taken into consideration. Under such circumstance, the relative permittivity of A decrease from 17.4 to 1.44 while the counterpart of B ranges from 0.003 to 0.77 as the radius *r* changes from *a* to *b*. Obviously, the material parameters for cloak with 20 layers approach the ideal case when compared to that of cloak with 10 layers. To demonstrate the performance of the cloak with different layers, the spatial distribution of magnetic field is simulated and plotted in Figure. 8. From Figure. 8(a) and (b), it can be found that the cloak with alternative AB layers from inside to outside has a better performance than that of BA layers while it is divided into 10 layers. Such a phenomenon can be also observed in Figure. 8(c) and (d). Besides, by comparing Figure. 8(a) and (c), it can be inferred that the cloak with more layers can better hiding the object inside it. That is reasonable since the impedance at the outer boundary is nearly the same as the impedance of the free space.

## IV. CONLCUSION

One kind of transformation functions with two parameters is proposed to realize the nonmagnetic invisibility cloak with minimized scattering cross section. Numerical simulation verifies the well performance of the cloaks. Moreover, total scattering cross section of the different cloaks is computed to illustrate the scattering characteristics of the cloaks. In addition, alternating layered system consisting of two isotropic materials is employed to realize the nonmagnetic invisible cloak from the prospective of practical realization

## ACKNOWLEDGEMENT

This work was supported in part by the State Key Program for Basic Research of China grants 2007CB613206 and 2006CB921704; the National Natural Science Foundation of China grants 10725418, 10734090, and 10990104; the Fund of

Shanghai Science and Technology Foundation grant 09DJ1400203, 09ZR1436100, 10JC1416100, and 10510704700.

**List of Figure Captions**

**FIG. 1:** (color online) Anisotropic material parameters $\varepsilon_r$ and $\varepsilon_\theta$ for different s and t (a) s=1, t=0.5, 1, 1.5 (b) t=1, s=0.5, 1, 1.5

**FIG. 2:** (color online) Computational domain and details for the full wave simulation. The space is surrounded by the PML. The incident wave propagates from right to left.

**FIG. 3:** (color online) The spatial distribution of magnetic field and scattering magnetic field. (a-b) magnetic field distribution for nonmagnetic cloak with s=1 and t=0.5, 2 (c-f) scattering magnetic field distribution for s=1 and t=0.5, 2

**FIG. 4:** (color online) The normalized scattering cross section of nonmagnetic invisibility cloaks for s=1 versus t

**FIG. 5:** (color online) The spatial distribution of magnetic field and scattering magnetic field. (a-b) magnetic field distribution for nonmagnetic cloak with t=1 and s=0.5, 2 (c-f) scattering magnetic field distribution for t=1 and s=0.5, 2

**FIG. 6:** (color online) The normalized scattering cross section of nonmagnetic invisibility cloaks for t=1 versus s

**FIG. 7:** (color online) The normalized scattering cross section of nonmagnetic invisibility cloaks for $0<s\leq2$ and $0<t\leq2$

**FIG. 8:** (color online) The magnetic field distribution of divisional cloak with 10 and 20 layer. Each layer is composed of two isotropic materials A and B. (a) layered AB for 10 layers. (b) layered BA for 10 layers (c) layered AB for 20 layers. (d) layered BA for 20 layers.

**FIG. 1**

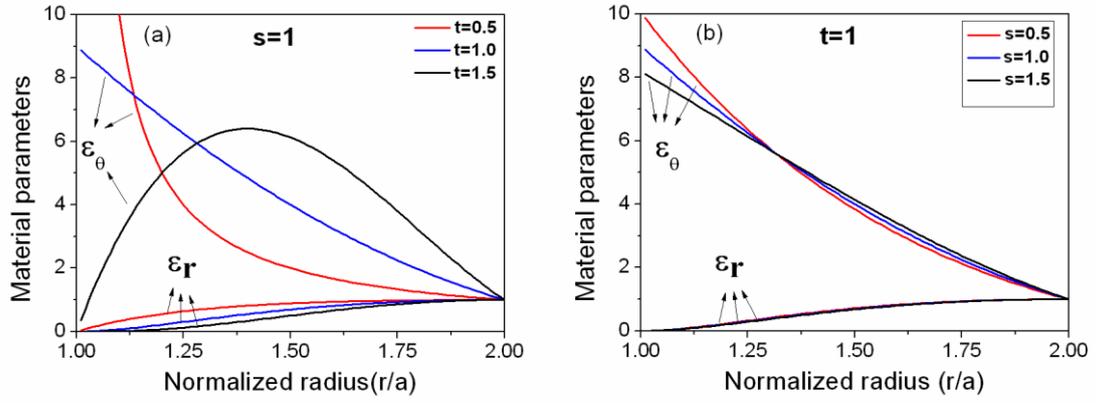

**FIG. 2**

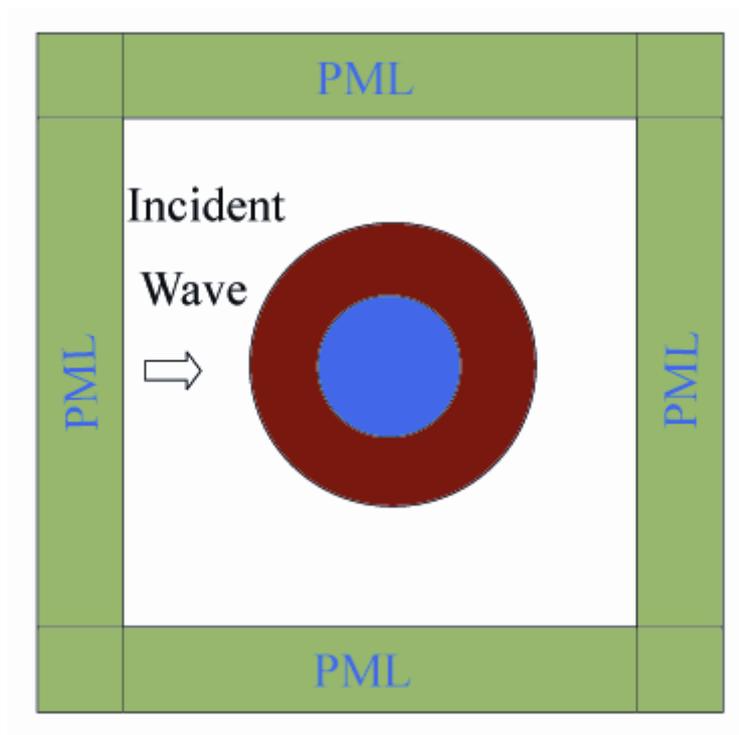

**FIG.3**

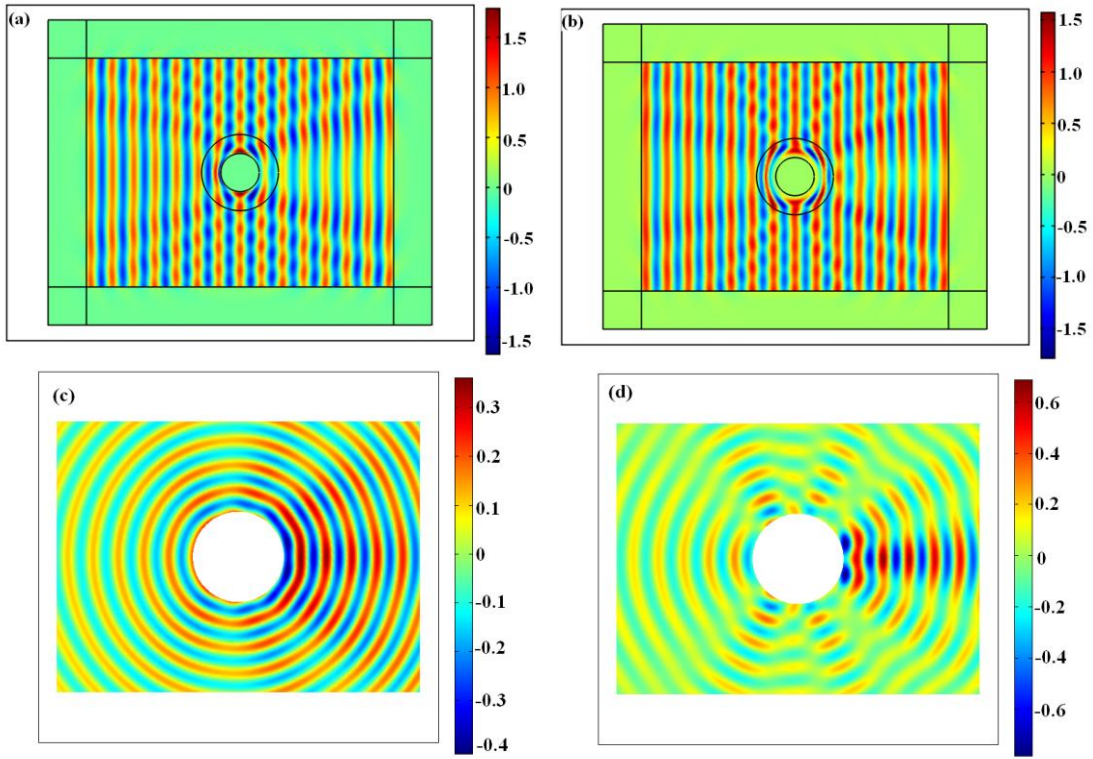

**FIG. 4**

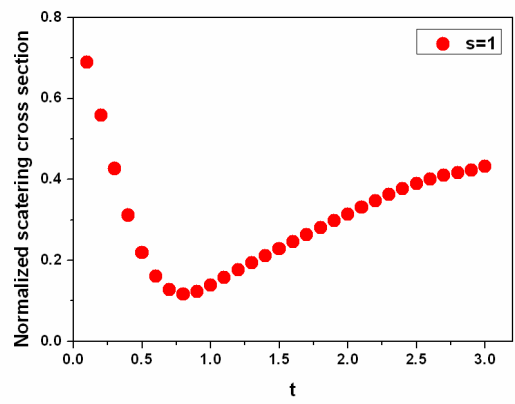

**FIG. 5**

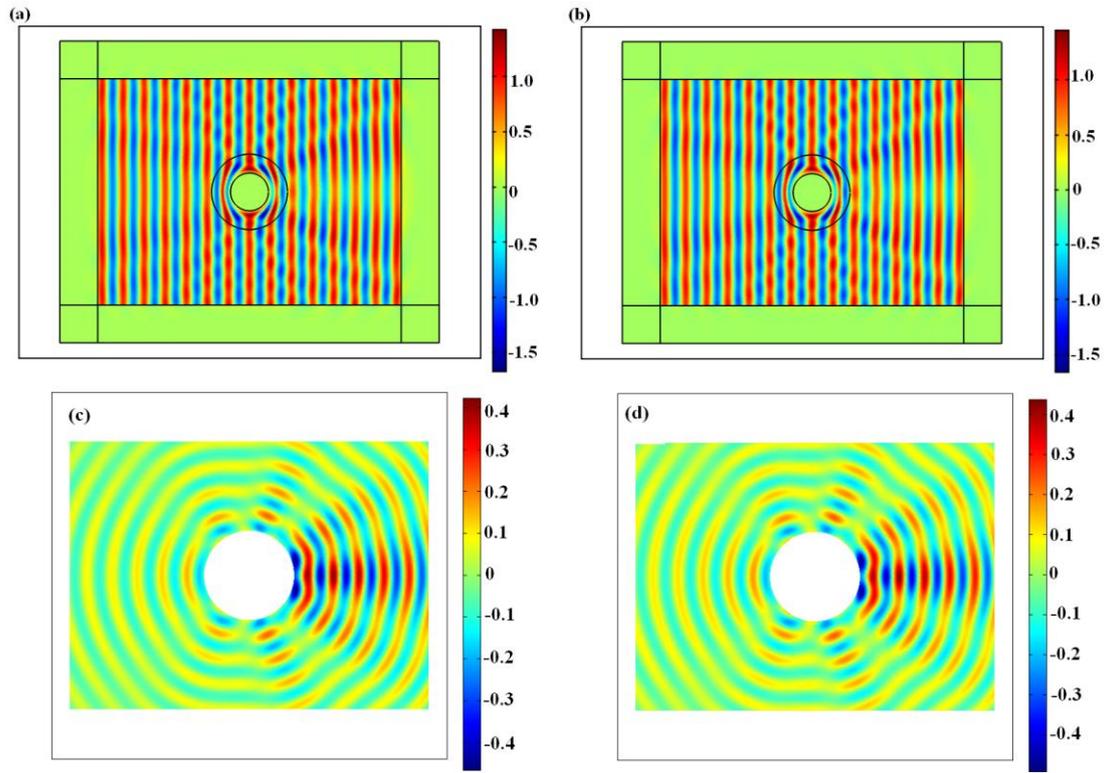

**FIG. 6**

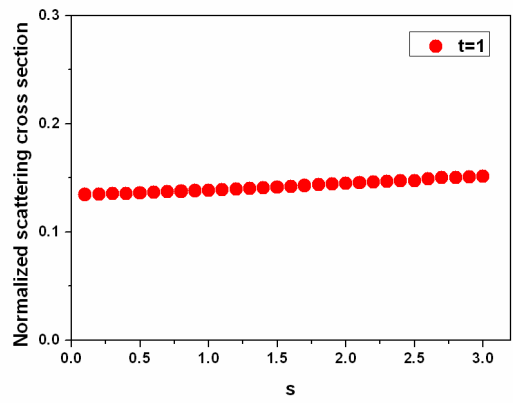

**FIG. 7**

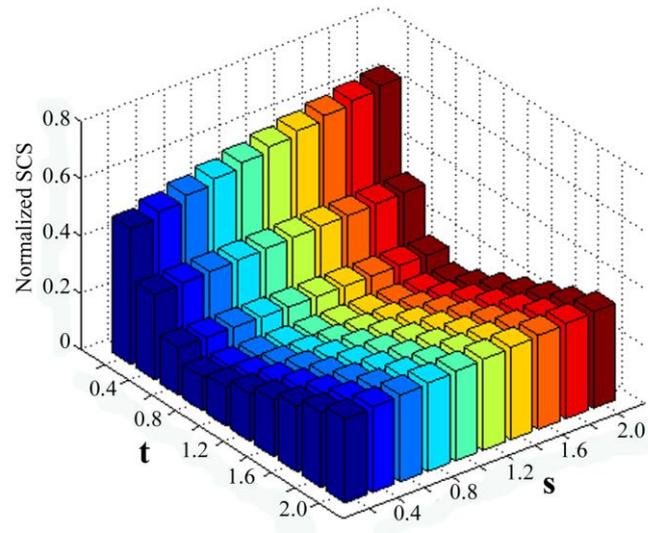

**FIG. 8**

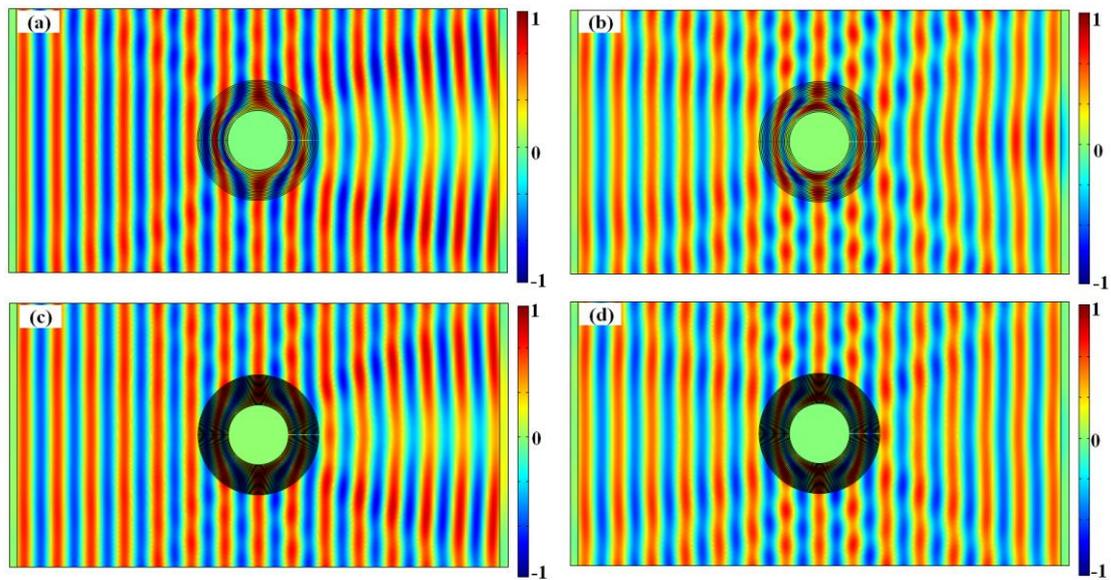